\documentclass[runningheads]{svmult}

\usepackage{makeidx}   
\usepackage{graphicx}  
\usepackage{subeqnar}  
\usepackage{multicol}  
\usepackage{physprbb}  
\makeindex             

\newcommand{\be}{\begin{equation}}
\newcommand{\ee}{\end{equation}}
\newcommand{\beann}{\begin{eqnarray*}}
\newcommand{\eeann}{\end{eqnarray*}}
\newcommand{\bea}{\begin{eqnarray}}
\newcommand{\eea}{\end{eqnarray}}
\newcommand{\lb}{\label}

\begin{document}
\title*{Is there an information-loss problem\protect\newline 
for black holes?}
\toctitle{Is there an information-loss problem
\protect\newline for black holes?}
%
%
\titlerunning{Information-loss problem}
%
\author{Claus Kiefer\inst{}}
\authorrunning{Claus Kiefer}
%
%
\institute{Institut f\"ur Theoretische Physik, Universit\"at zu K\"oln,
Z\"ulpicher Str.~77, 50937~K\"oln, Germany.}

\maketitle              

\begin{abstract}
Black holes emit thermal radiation (Hawking effect). If after black-hole
evaporation nothing else were left, an arbitrary initial state
would evolve into a thermal state (`information-loss problem').
Here it is argued that the whole evolution is unitary and that the
thermal nature of Hawking radiation emerges solely through decoherence --
the irreversible interaction with further degrees of freedom. 
For this purpose a detailed comparison with an analogous case in
cosmology (entropy of primordial fluctuations) is presented.
Some remarks on the possible origin of black-hole entropy 
due to interaction with other degrees of freedom are added. This might
concern the interaction with quasi-normal modes or with background fields
in string theory. 
\end{abstract}

\section{The information-loss problem}
Black holes are amazing objects. According to general relativity,
stationary black holes are fully characterised by just three numbers:
Mass, angular momentum, and electric charge. This ``no-hair theorem''
holds within the Einstein-Maxwell theory in four spacetime dimensions.
It reminds one at the properties of a macroscopic gas which can be described
by only few variables such as energy, entropy, and pressure. 
In fact, there exist laws of black-hole mechanics which are analogous to
the laws of thermodynamics (see e.g. \cite{CK99} for a detailed review).
The temperature is proportional to the surface gravity, $\kappa$,
of the black hole, and the entropy is proportional to its area,
$A$. That this correspondence is not only a formal one, but possesses
physical significance, was shown by Hawking in his seminal paper
\cite{SH75}. Considering quantum field theory on the background
of a collapsing star (see Fig.~1), it is found that the black hole
radiates with a temperature proportional to $\hbar$,
\be
T_{\rm BH}=\frac{\hbar\kappa}{2\pi k_{\rm B}}\ .
\ee
\begin{figure}[b]
\begin{center}
\includegraphics[width=.5\textwidth]{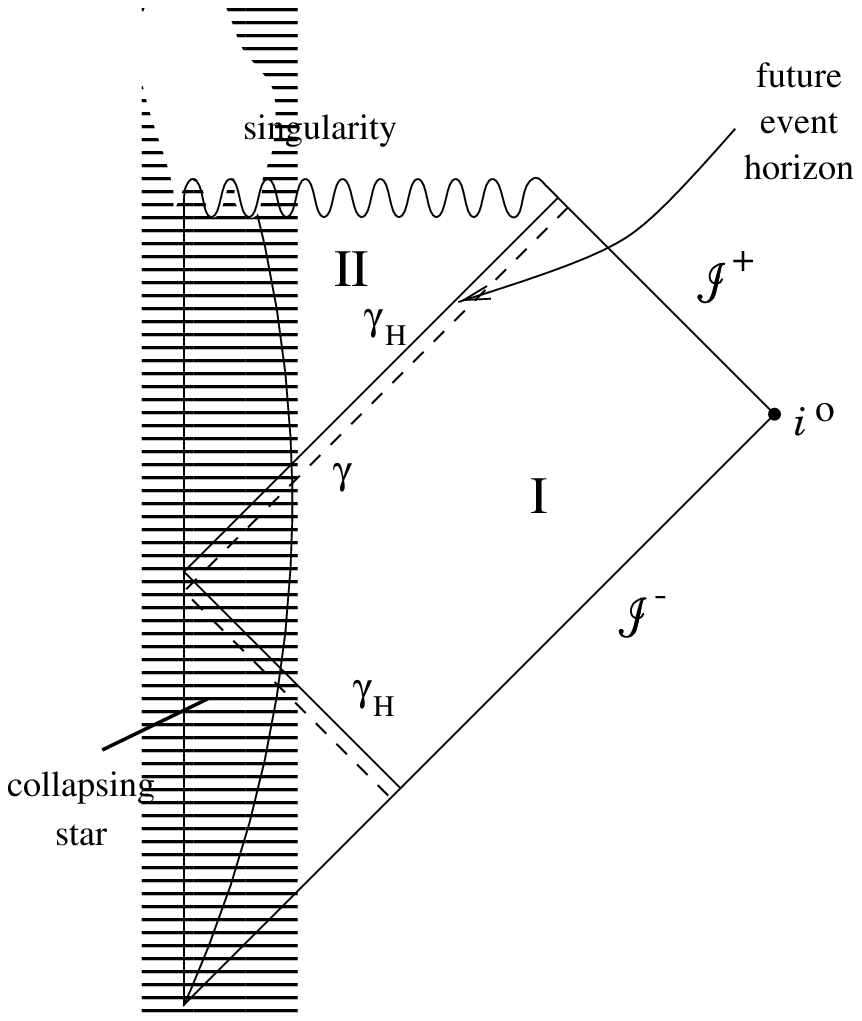}
\end{center}
\caption[]{Penrose diagram of a star to form a black hole}
\label{eps1}
\end{figure}
The origin of this temperature is the presence of a horizon. 
Due to the high gravitational redshift in its vicinity
(symbolised in Fig.~1 by the dashed line $\gamma$ near the 
horizon $\gamma_{\rm H}$), the vacuum modes
are excited for a very long time, until the black hole has evaporated. 
The entropy connected with this temperature is given by 
the `Bekenstein-Hawking' formula,
\be
\lb{SBH}
S_{\rm BH}=\frac{k_{\rm B}A}{4G\hbar}\ .
\ee
In the case of a spherically-symmetric (`Schwarzschild') black hole, one has
$\kappa=(4GM)^{-1}$, and the Hawking temperature is given by
\be
\lb{TH}
T_\mathrm{BH}= \frac{\hbar}
  {8 \pi G k_\mathrm{B} M} \,\approx   6.2\times 10^{-8}
  \frac{M_\odot}{M}\ \mathrm K\ .
\ee
If the quantum field on the black-hole background is a massless scalar field,
the expectation value of the particle number for a mode with wave number
${\mathbf k}$ is given by
\be
\lb{nk}
\langle n_k\rangle=\frac{1}{{\rm e}^{8\pi\omega GM}-1}\ ,
\ee
with $\omega=\vert{\mathbf k}\vert=k$. This is a Planck distribution with
temperature (\ref{TH}). The usual interpretation is `particle creation':
a mode with pure positive frequency will evolve 
(along the dashed line in Fig.~1) into a superposition of positive
and negative frequences. Therefore, an initial vacuum will evolve into
a superposition of excited states (`particles'). In the Heisenberg
picture, this is described by the occurrence of a non-vanishing
Bogolubov coefficient $\beta$ \cite{CK99,SH75}.

An alternative point of view arises through the use of the
Schr\"odinger picture. Taking the scalar field in Fourier space,
$\phi(k)$, the initial vacuum state can be expressed as a Gaussian wave
functional. One assumes for simplicity that the full wave functional
can be written as a product over independent modes, 
$\Psi=\prod_k\psi_k$ (one can imagine putting the whole system into
a box). Then,
\be
\psi_k\propto \exp\left[-k|\phi(k)|^2\right]\ ,
\ee
see e.g. \cite{Jackiw} for details about the functional Schr\"odinger
picture. With this initial state, the functional Schr\"odinger equation
can be solved exactly \cite{DK}. Due to the dynamical
gravitational background,
the various $\psi_k$, albeit always of Gaussian form, develop a
complex function in the exponent \cite{DK,CK01},
\be
\lb{psik}
\psi_{k}\propto \exp\left[-k\ {\rm coth}(2\pi kGM+{\rm i}kt)
 \vert\phi(k)\vert^2\right]\ .
\ee
This is still a {\em pure state}, but the expectation value of the
particle number operator with respect to this state is of the same
Planckian form as (\ref{nk}),
\be
\lb{nkpsik}
 \langle\psi_k|n_k|\psi_k\rangle=
\frac{1}{{\rm e}^{8\pi\omega GM}-1}\ .
\ee
As a side remark I note that such a result can also be obtained
from the wave functional solving the Wheeler-DeWitt equation in the
WKB approximation \cite{VKSW}. 
A state such as (\ref{psik}) is well known from quantum optics and called
a {\em two-mode squeezed state}. That Hawking radiation can be described
in this terminology was first recognised -- using the Heisenberg picture --
in \cite{GS}.

For spatial hypersurfaces that enter the horizon one must trace out the
degrees of freedom which reside inside the horizon. This results in a
thermal density matrix in the outside region \cite{Israel},
with the temperature being equal to the Hawking temperature
(\ref{TH}). It is a general property of two-modes squeezed states that
one gets a thermal density matrix if half of the modes is being
traced out \cite{squeezed}. The emergence of a density matrix is not
surprising. Taking into account only part of the degrees of freedom
one is dealing with an open quantum system \cite{deco}. Such systems
do not obey a unitary dynamics, but are described by master equations.
This is not related to any information-loss in the full system
(including the interior of the black hole), but only to an effective
information loss (or gain) for the reduced system.

The black-hole entropy (\ref{SBH}) is much bigger than the entropy
of a collapsing star. The entropy of the Sun, for example, is
$S_{\odot}\approx 10^{57}$, but the entropy of a solar-mass black hole
is $S_{\rm BH}\approx 10^{77}$, i.e. twenty orders of magnitudes larger
(all entropies are measured in units of $k_{\rm B}$). If all matter in the
observable Universe were in a single gigantic black hole, its 
entropy would be $S_{\rm BH}\approx 10^{123}$. Black holes thus seem to be
the most efficient objects for swallowing information. 

Due to Hawking radiation, black holes have a finite lifetime.
It is given by
\be
\lb{tauBH}
 \tau_\mathrm{BH} \approx \left(\frac{M_0}{m_\mathrm{p}} \right)^3
  t_\mathrm{p} \approx 10^{65} \left(\frac{M_0}{M_\odot} \right)^3
  \mathrm{years}\ ,
\ee
where $m_\mathrm{p}$ and $t_\mathrm{p}$ denote Planck mass and Planck
time, respectively. The question now arises what happens at the end of
black-hole evaporation. If only thermal radiation 
were left behind, an arbitrary initial state (for the star collapsing
to form a black hole) would evolve into a mixed state. This process
does not happen in the standard quantum theory for closed systems. 
There, the entropy
\be
\lb{Entropy}
S=-k_{\rm B}{\rm Tr}(\rho\ln\rho)
\ee
is conserved for the full system.
Since a thermal state contains least information, one would be faced
with the {\em information-loss problem}. This is, in fact, what Hawking
speculated to happen. The calculations in \cite{SH75} are, however,
restricted to the semiclassical approximation (gravity classical and
matter quantum), which breaks down when the black hole approaches the
Planck mass. The final answer can only be obtained within quantum gravity.
The options are \cite{Page}
\begin{itemize}
\item Information is indeed {\em lost} during the evaporation,
i.e. the evolution is non-unitary,
 \begin{displaymath}
\rho\to \$\rho\neq S\rho S^{\dagger}\ .
\end{displaymath}
\item The full evolution is {\em unitary}, but this cannot be seen
in the semiclassical approximation.
\item The black hole leaves a {\em remnant} carrying all the
information.
\end{itemize}
The state (\ref{psik}) also holds on a spatial surface that in Fig.~1 
would start at the intersection of the collapsing star with the horizon
and extend to spatial infinity $i^0$, i.e. a surface that does not enter
the horizon. In fact, such surfaces seem quite natural, since they
correspond to constant Schwarzschild time far away from the black hole.
The question then arises where the thermal  
nature of Hawking radiation comes from; 
although (\ref{nkpsik}) is Planckian, the state (\ref{psik}) is pure
and the differences to a thermal distribution can be recognised in
higher-order correlation functions. 

I shall argue in Sect.~3 that the thermal appearance of Hawking radiation
can be understood, even for the pure state (\ref{psik}), through
{\em decoherence} -- the irreversible and unavoidable interaction
with the environment \cite{deco}. For this purpose it will be appropriate
to rewrite (\ref{psik}) in a form where the two-mode squeezed nature becomes
explicit. In fact, one can rewrite $\psi_k$ in the form
\bea
\lb{squeezed}
\psi_{k} &\propto&
\exp\left[-k\frac{1+{\rm e}^{2{\rm i}\varphi_k}\tanh r_k}
 {1-{\rm e}^{2{\rm i}\varphi_k}\tanh r_k}\vert\phi(k)\vert^2\right]\nonumber\\
&\equiv& \exp\left[-(\Omega_R+{\rm i}\Omega_I)\vert\phi(k)\vert^2\right]\ ,
\eea
where the {\em squeezing parameter} $r_k$ is given by
\be
\lb{rk}
\tanh r_k=\exp(-4\pi\omega GM)\ ,
\ee
and the {\em squeezing angle} $\varphi_k$ reads
\be
\lb{phik}
\varphi_k=-kt\ .
\ee
Thus, $r_k\to 0$ for $k\to\infty$ and
$r_k\to\infty$ for $k\to 0$: modes with bigger wavelength are more
squeezed than modes with smaller wavelength. 
At the maximum of the Planck spectrum one has $r\approx 0.25$.
This corresponds to $\langle n_k\rangle=\sinh^2r_k\approx 0.06$ for
the expectation value of the particle number.

At $kt=0$ the squeezing is in $\phi$, whereas at
$kt=\pi/2$ the squeezing is in $p_{\phi}$, the momentum conjugate to $\phi$.
The ratio of the corresponding widths is $\tanh^2(2\pi kGM)$
($\approx 0.37$ at the maximum of the Planck spectrum).
Before I apply this to the study of decoherence, it is appropriate to
review the analogous situation in inflationary cosmology. This will serve
to understand the similarities to and the differences from the
black-hole case. 

\section{Entropy of cosmological fluctuations}

One of the most important advantages of an inflationary
scenario for the early Universe is the possibility to
obtain a dynamical explanation of structure formation
(see e.g. \cite{LL} for a review). Quantum vacuum fluctuations
are amplified by inflation, leading to a squeezed state
(corresponding to particle creation in the Heisenberg picture).
The scalar field(s) and the scalar part of the metric
describe primordial density fluctuations, while the tensor
part of the metric describes gravitons. The generation of
gravitons is, in fact, an effect of linear quantum gravity.
The primordial fluctuations can at a later stage serve as
seeds for structure (galaxies and clusters of galaxies).
They exhibit themselves in the CMB temperature anisotropy
spectrum. The gravitons may lead to a stochastic
gravitational-wave background that might in principle be
observable with space-based experiments.

An important issue in the theoretical understanding
of the above process is the exact way in which these
quantum fluctuations\footnote{In the following,
``fluctuations'' refers to primordial density
fluctuations as well as gravitons.} become classical
stochastic variables, see e.g. \cite{KP} for discussion and references.
Two ingredients are
responsible for the emergence of their classical behaviour.
Firstly, inflation leads to a huge squeezing of the quantum
state for the fluctuations and, therefore, to a huge
particle creation (the number of generated particles
for a mode with wave number $k$ is
$N_k=\sinh^2r_k$, where $r_k$ is the
squeezing parameter.) But in the limit of large $r_k$
the quantum state becomes an approximate WKB state,
corresponding in the Heisenberg picture to neglecting
the part of the solution which goes as ${\rm e}^{-r_k}$ and
which is called the ``decaying mode''. Secondly, interactions
with other, ``environmental'', degrees of freedom lead to
the field-amplitude basis (here called $y_k$) as
the classical basis (``pointer basis'') with respect to
which interferences become unobservable.
This process of {\em decoherence} transforms the $y_k$ into effective classical
stochastic quantities. On the largest cosmological scales one finds for
the squeezing parameter $r_k\approx 100$, far beyond any values which
can be attained in the laboratory. 

Whereas the quantum theory therefore does not lead to
deviations from the usual predictions (based on a 
phenomenological classical stochastic theory) of the inflationary
scenario, the entropy of the fluctuations depends on their
quantum nature (on the presence of the decaying mode, albeit small),
see \cite{KPS1}.
The entropy of the squeezed quantum state is of course zero,
because it is a pure state. Due to the interaction with other
degrees of freedom, however, the fluctuations have to
be described by a density matrix $\rho$. The relevant
quantity is then the von~Neumann entropy (\ref{Entropy}).

In the context of primordial fluctuations, different mechanisms of
coarse-graining have been investigated in order to
calculate the local entropy. It was found
that the maximal value for the entropy is
$S_{\rm max}=2r_k$, resulting from a coarse-graining
with respect to the particle-number basis, i.e. 
non-diagonal elements of the reduced density matrix $\rho$
are neglected in this basis. Consideration of the corresponding
contour of the Wigner ellipse in phase space shows that this
would smear out the thin elongated ellipse of the squeezed
state (corresponding to the high values of $r_k$)
into a big circle. Does such a coarse-graining
reflect the actual process happening during inflation?
Since the pointer basis in the quantum-to-classical transition
is the field-amplitude basis and {\em not} the particle-number
basis (which mixes the field variable with its canonical
momentum), one would expect that coarse-graining should be
done with respect to $y(k)$. This would then lead to $S=S_{\rm max}/2$,
which is noticeably different from maximal entropy \cite{KPS1}. 

A crucial observation for the calculation of (\ref{Entropy})
is to note that the wavelength of the amplified
fluctuations during inflation is bigger than the
horizon scale, i.e bigger than $H_{\rm I}^{-1}$, where $H_{\rm I}$
is the Hubble parameter of inflation (here taken to be
approximately constant for simplicity). This prevents a direct
causal interaction with the other, environmental, fields.
However, nonlocal quantum correlations can still develop
due to interaction terms in the total Hamiltonian.
Since, as remarked above, the interaction is local in
$y(k)$ (as opposed to its momentum), the density
matrix will be of the form (suppressing $k$ for simplicity)
\begin{equation}
\rho_{\xi}(y,y')=\rho_0(y,y')
\exp\left(-\frac{\xi}{2}(y-y')^2\right)\ ,
\label{rhoxi}
\end{equation}
where $\rho_0(y,y')$ denotes the density matrix
referring to the squeezed state, which is a Gaussian state,
see e.g. \cite{deco} for a discussion. Eq.~(\ref{rhoxi})
guarantees that interferences in the $y$-basis 
become small, while the probabilities (diagonal elements)
remain unchanged. The details of the interaction are
encoded in the phenomenological parameter $\xi$ and are
not needed for our discussion. Decoherence is efficient
if the Gaussian in (\ref{rhoxi}) dominates over the Gaussian
from the squeezed state described by $\rho_0$. This leads to
the condition \cite{KPS1}
\begin{equation} 
\frac{\xi {\rm e}^{2r}}{k}\gg 1 \ , \label{dec}
\end{equation}
which is called the {\em decoherence condition}.

The entropy of the fluctuations is now calculated by
setting $\rho=\rho_{\xi}$ in (\ref{Entropy}).
For an arbitrary Gaussian density
matrix, the result has been obtained in
\cite{JZ}, see also Appendix~A2.3 in \cite{deco}.
With the abbreviation 
\begin{equation}
\chi=\frac{\xi}{k}(1+4\sinh^2r)\ ,
\end{equation}
the result is in our case given by the expression
\begin{equation}
S=-\ln 2+\frac{1}{2}\ln\chi-
 \frac{\sqrt{1+\chi}}{2}\ln\frac{\sqrt{1+\chi}-1}
 {\sqrt{1+\chi}+1} \ .
\end{equation}
The decoherence condition means $\chi\gg1$, which leads for arbitrary $r$ to
\begin{equation}
S\approx 1+\frac{1}{2}\ln\frac{\xi(1+4N)}{4k}\ .
\end{equation} 
In the high-squeezing limit, ${\rm e}^r\to\infty$, this yields
\begin{equation}
S\approx 1-\ln2+\frac{1}{2}\ln\frac{{\rm e}^{2r}\xi}{k}=1+
\frac{1}{2}\ln \frac{N\xi}{k}~. \label{entropy2}
\end{equation}
In phase space this corresponds to
$S\approx\ln A$, where $A$ is the area of the Wigner ellipse. 
Application of the decoherence condition (\ref{dec}) leads to
\begin{equation}
S\gg 1-\ln2\approx 0.31\ , 
\end{equation}
where $\gg$ holds here in a logarithmic sense (it directly holds
for the number of states ${\rm e}^S$). Therefore, decoherence
already occurs after few bits of information are lost.
This is much less than the maximal entropy, which is
obtained if the ellipse is smeared out to a big circle,
corresponding to the choice $\xi/k={\rm e}^{2r}$ in (\ref{entropy2}),
and leading to $S_{\rm max}=2r$, as has been remarked above.
That only few bits of information loss can be sufficient for
decoherence is well known from quantum optics \cite{deco}.
The correlation between $y$ and
its canonical momentum is only preserved
if $S\ll S_{\rm max}=2r$ because otherwise the squeezed Wigner
ellipse is no longer recognisable.

One would expect that the maximal possible entropy due to
quantum entanglement alone (i.e. without dynamical back reaction)
is obtained if the coarse-graining
is performed exactly with respect
to the {\em field-amplitude basis} $y$. As remarked above,
this would lead to
$S=r$. Inspecting (\ref{entropy2}), this would correspond to the
choice $\xi=k$. Therefore, as long as the modes are outside the
horizon, one would expect to have $\xi<k$ which has a
very intuitive interpretation: the coherence length $\xi^{-1/2}$
is larger than the width of the ground state ($r=0$),
so that the environment does not spoil the property of the
quantum state being squeezed in some direction compared to the
ground state. It can be shown that the correlation between
$y$ and the conjugate momentum remains for a sufficiently long time after the
second horizon crossing (in the postinflationary phase), so that
it really leads to the observed acoustic peaks (B-polarisation
for gravitational waves) in the CMB \cite{KLPS}.
The information contained in these peaks can be interpreted
as a measure for the deviation of the entropy from the
maximal entropy. Therefore, coarse-grainings that lead to
maximal entropy would prevent the occurrence of such peaks
and would thus be {\em in conflict with observation}.

This analysis has also borne out an interesting analogy of the primordial 
fluctuations with a chaotic system: the Hubble parameter
corresponds to a Lyapunov exponent, although our
system is not chaotic, but only classically unstable \cite{KPS1}.
In the next section the comparison of the cosmological case with the
black-hole case will be made.

\section{Hawking radiation from decoherence}

As in the cosmological case, the quantum state corresponding to
Hawking radiation is a two-mode squeezed state. There are, however, pronounced
differences in the black-hole case. As has been remarked at the end of
Sect.~1, the squeezing parameter for the maximum of the Planck distribution
is only $r_k\approx 0.25$, which is far below the values attained
from inflation. High squeezing values are only obtained for very big
wavelengths. 

The quantum state can again be represented by the contour of the
Winger ellipse in phase space. In the cosmological case the rotation of this
ellipse is very slow, the corresponding time being about the age of the
Universe \cite{KLPS}. This reflects the fact that it is not allowed to
coarse-grain this ellipse into a big circle (Sect.~2).
What about the situation for black holes?
Again, the Wigner ellipse rotates around the origin, and
the typical timescale is given by
\be
\lb{tk}
t_k=\frac{\pi}{2k}\ .
\ee
This corresponds to the exchange of squeezing between $\phi$ and 
its conjugate momentum $p_{\phi}$, cf.
the end of Sect.~1. Evaluated at the maximum of the Planck spectrum, one has
\be
t_k({\rm max})\approx 14GM\approx 7\times 10^{-5}\frac{M}{M_{\odot}}\ {\rm s}\ , 
\lb{time}
\ee
which is much smaller than typical observation times. It is for this 
reason that a coarse-graining with respect to the squeezing angle
{\em can} be performed. Squeezed states are extremely sensitive to interactions
with environmental degrees of freedom \cite{deco}.
 In the present case of a quickly
rotating squeezing angle this interaction leads to a diagonalisation
of the reduced density matrix with respect to the particle-number basis
\cite{Pro}, not the field-amplitude basis.
 Thereby the local entropy is maximised, corresponding
to the coarse-graining of the Wigner ellipse into a circle.
The value of this entropy can be calculated along the lines of
Sect.~2. In contrast to the cosmological case one finds the standard
expression for a thermal ensemble,
\be
S_k=(1+n_k)\ln(1+n_k)-n_k\ln n_k \stackrel{r_k\gg 1}{\longrightarrow}
2r_k \ . \lb{entropy3}
\ee
The integration over all modes gives $S=(2\pi^2/45)T_{\rm BH}^3V$, which is
just the entropy of the Hawking radiation with temperature
$T_{\rm BH}=(8\pi GM)^{-1}$. In this way, the pure squeezed state
becomes indistinguishable from a canonical ensemble with temperature
$T_{\rm BH}$ \cite{CK01}. 

Independent of this practical indistinguishability
from a thermal ensemble, the state remains
a pure state. In fact, for timescales smaller than $t_k$ the
above coarse-graining is not allowed and the difference to a thermal state
could be seen in principle. For the case of a primordial black hole
with mass $M\approx 5\times 10^{14}\ {\rm g}$ one has at the maximum of the
Planck spectrum $t_k({\rm max})\approx 1.7\times 10^{-23}\ {\rm s}$.
The observation of Hawking radiation at smaller times could then reveal the
difference between the (true) pure state and a thermal state.  
It is also clear from (\ref{tk}) that large wavelengths (much larger than
the wavelength corresponding to the maximum of the Planck spectrum)
have a much longer rotation time for the Wigner ellipse.
This would also offer, in principle, the possibility to distinguish
observationally between pure and mixed states -- provided, of course,
that primordial black holes exist and can be observed.

To summarise, no mixed state for the total system has appeared 
at any stage of this discussion. This indicates that the full quantum
evolution of collapsing star plus scalar field evolves unitarily. 
The thermal nature of Hawking radiation thus only emerges through
coarse-graining. For hypersurfaces entering the horizon, this is achieved
by tracing out degrees of freedom referring to the interior
\cite{Israel}. As has been emphasised here, however, a similar result
holds for hypersurfaces that stay outside the horizon. Such hypersurfaces
are a natural choice for asymptotic observers. 
The mixed appearance of the pure state for the quantum field
is due to its squeezed nature. Squeezed states are very sensitive to
interactions with other fields, even for very weak coupling.
Such interactions lead to decoherence for the Hawking radiation.
The thermal nature of the reduced density matrix is a consequence
of the presence of the horizon as encoded in the particular
squeezed state (\ref{psik}), cf.~\cite{DK}. 

Similar conclusions hold 
for hypersurfaces entering the horizon \cite{CK01}. A general problem is,
however, that the evolution along different foliations is in general not
unitarily equivalent \cite{TV}. Observations at spatial infinity
cannot thus be correctly recovered from an arbitrary foliation.
The precise relationship between observation and choice of hypersurfaces
is not yet properly understood and deserves investigation. 

Since no information loss occurs in the first place, there does not seem
to be any case for an information-loss problem. The above line of argument
holds, of course, only within the semiclassical regime, neglecting
quantum effects of the gravitational field itself. 
Only a full quantum theory of gravity can give an exact description of
black-hole evaporation. One would, however, not expect that
a unitary evolution during the semiclassical phase would 
be followed by a sudden information loss at the final stage.

\section{Bekenstein-Hawking entropy through decoherence?}

The above discussion was concerned with Hawking radiation. I have argued that
its mixed appearance is due to the decohering influence of other fields.
The entropy of the Hawking radiation, Eq. (\ref{entropy3}), is the
result of a coarse-graining and application of von~Neumann's
formula (\ref{Entropy}). But what about the entropy of the black hole
itself? In other words, can the Bekenstein-Hawking formula (\ref{SBH})
be recovered, within the semiclassical approximation, along the same
lines?

For an answer, one should be able to specify microscopic degrees of
freedom of the gravitational field itself. This has been achieved,
in certain situations, within tentative approaches to quantum
gravity: loop quantum gravity (see e.g. \cite{ash}) and string theory
(see e.g. \cite{GH98}). In string theory, the derivation of
(\ref{SBH}) in achieved in an indirect way: using duality and the properties
of `BPS states', a black hole in the limit of strong coupling
($g\gg 1$) can be related to a bound collection of `D-branes' 
in flat space in the
limit of weak coupling ($g\ll1$). Both configurations should have the same
number of quantum states -- this is the property of BPS states.
 For the D-branes, standard formulas of string theory
give a definite answer which coincides with the expression
(\ref{SBH}) for the duality-related black hole.  

It was argued in \cite{RM97} that black holes are inherently associated
with mixed states, and that pure D-brane states rapidly develop
entanglement with other degrees of freedom, leading to decoherence.
Only the decohered D-branes should be associated with black holes.
In fact, the derivation of (\ref{SBH}) employs, at least implicitly,
the use of decohered D-branes.

A related situation was investigated in \cite{AR99}.
There, the decay of a massive string state (i.e. a string state with high
excitation number $n$) and small coupling was addressed. 
The mass is given by
\be
M^2=\frac{n}{l_{\rm s}^2} \ , \quad n\gg1\ ,
\ee
where $l_{\rm s}$ is the string length. The degeneracy $N_n$ of a level $n$
is given by
\be
N_n\sim {\rm e}^{a\sqrt{n}}={\rm e}^{aMl_{\rm s}}
\ , \quad a=2\pi\sqrt{\frac{D-2}{6}}\ ,
\ee
where $D$ is the number of spacetime dimensions
in which the string moves. The decay spectrum of
a single excited state does not exhibit any thermal properties. 
However, if one averages over all the degenerate states with the same mass
$M$, the decay spectrum is of Planckian form, with the temperature
given by the Hagedorn temperature $T_{\rm H}=(al_{\rm s})^{-1}$
\cite{AR99}. It is speculated that the reason is decoherence
generated by the entanglement with quantum background fields 
being present in the string spectrum. 

It might well be that (\ref{SBH}) can generally be justified in an 
analogous manner,
either in loop quantum gravity or string theory. Quantitative calculations
in this direction have still to be performed. They should in particular
reveal the universal nature of the Bekenstein-Hawking entropy. 
One might expect that therein the {\em quasi-normal modes} of a black hole
could play a crucial role in serving as an environment. For the Schwarzschild
black hole the maximal value for the real part of the mode frequency 
(energy) is \cite{HM}
\be
\lb{qnm}
\hbar \omega =\frac{\hbar\ln 3}{8\pi GM}=(\ln 3)k_{\rm B} T_{\rm BH}\ ,
\ee
This frequency seems to play a crucial role in
the calculation of (\ref{SBH}) from loop quantum gravity \cite{qnm}.
In quantum cosmology, the global structure of spacetime assumes
classical properties through interaction with higher modes
\cite{qc,deco}. In a similar way one could envisage the quasi-normal modes
to produce a classical behaviour for black holes -- the entropy
(\ref{SBH}) would then result as the entanglement entropy of
the correlated state between black hole and quasi-normal modes. Since in the
corresponding calculation a sum over all modes has to be performed,
one might expect the maximal frequency (\ref{qnm}) to play a crucial role.
The details, however, are far from being explored.
To understand black-hole entropy as an entanglement entropy has been
tried before, see e.g. \cite{honnef} and the references therein.
A special example is induced gravity where (\ref{SBH}) was recovered
in the presence of non-minimally coupled fields \cite{FFZ}. 
(This result might be of relevance to string theory.) Here it is
suggested that the role of the environment is played by the
quasi-normal modes, which are (for large mode number) characteristic
of the black hole itself and therefore should be able to yield a universal
result.

It is known that spacetime
as such is a classical concept, arising from decoherence in
quantum gravity (see Sect.~5 in \cite{deco}). The event horizon of a black
hole is a spacetime concept and should therefore have no fundamental
meaning in quantum gravity. It should arise 
from decoherence in the semiclassical limit,
together with (\ref{SBH}).


%

\end{document}